\documentclass[letterpaper,pra,twocolumn,showpacs]{revtex4}
\usepackage{graphicx}
\usepackage{amsmath}
\usepackage{amssymb}

\begin{document}

\title{Quantum limits in image processing}

\author{V. Delaubert$^{1,2}$, N. Treps$^{1}$, C. Fabre$^{1}$, H.A. Bachor$^{2}$, and P. R\'efr\'egier$^{3}$ }
\affiliation{$^{1}$Laboratoire Kastler Brossel, Universit\'e Pierre
et Marie Curie-Paris6, CC74, 4 Place Jussieu, 75252 Paris cedex 05,
France} \affiliation{$^{2}$ARC Centre of Excellence for Quantum-Atom
Optics, Australian National University, Canberra, ACT 0200,
Australia} \affiliation{$^{3}$Physics and Image processing group,
Fresnel Institute,UMR 6133, Domaine Universitaire de
Saint-J\'er\^ome, 13397 Marseille cedex 20, France}

\date{\today}

\begin{abstract}
We determine the bound to the maximum achievable sensitivity in the
estimation of a scalar parameter from the information contained in
an optical image in the presence of quantum noise. This limit, based
on the Cramer-Rao bound, is valid for any image processing protocol.
It is calculated both in the case of a shot noise limited image and
of a non-classical illumination. We also give practical experimental
implementations allowing us to reach this absolute limit.
\end{abstract}


\maketitle

{\it Introduction.} - Many of the most sensitive techniques for the
measurement of a physical parameter - which we will call $p$ in the
following and assume to be a scalar - are optical. In some cases,
the total intensity or amplitude of the light beam varies with $p$
and conveys the information. It is then well known
\cite{Kimble,Fabre,Gao,SQL} that there exists a standard quantum
limit in the sensitivity of the measurement of $p$ when the light
beam is in a coherent state, and that it is possible to go beyond
this limit using sub-Poissonian or squeezed light. In other cases,
the parameter $p$ of interest only modifies the distribution of
light in the transverse plane and not its total intensity. The
present paper deals with this latter situation. For example, the
parameter $p$ modifies the position or direction of a light beam.
This configuration has been studied at the quantum level, both
theoretically and experimentally using split detectors
\cite{displacement11, displacement12,displacement13} or homodyne
detection \cite{Delaubert2006,displacement21}. But in many instances
the parameter $p$ affects in a complicated way the field
distribution in the detection plane (that we will call here the
image). For example a fluorescent nano-object imbedded in a
biological environment modifies the image recorded through a
microscope in a complex way because of diffraction. Nevertheless its
position can be determined from the information contained in the
image with a sensitivity which can be much better than the
wavelength \cite{nano}. In order to extract the parameter value in
such experiments, one needs to use detector arrays or CCD cameras
and to combine in an appropriate way the information recorded on the
different pixels.

When all the sources of technical noise have been removed in the
apparatus, quantum fluctuations still affect the optical measurement
and limit its sensitivity, in a way that can be readily calculated
for each specific measurement protocol. The purpose of this paper is
much broader. It is to answer the following question: what is the
lowest limit imposed by quantum noise to the accuracy of the
determination of $p$, independently of the information processing
protocol used for the extraction of information? As we will see,
this optimum limit depends only on the statistics of the
fluctuations of the incoming light. We use an approach based on the
Cramer-Rao Bound. This tool, widely used in the signal processing
community \cite{Refregier2002}, has already been applied to
different domains, such as gravitational wave detection
\cite{Nicholson} or diamagnetism \cite{Curilef}.

{\it Notations and assumptions.} - The parameter $p$ is measured
relative to an {\it a priori} value chosen for simplicity to be $0$.
Because of the quantum fluctuations in the optical measurements,
there will be an uncertainty on its estimation. An evaluation of
this uncertainty thus provides the precision on the determination of
the parameter $p$ around a zero value.

The mean value of the local complex electric field operator in the
image plane, normalized to a number of photons, will be written
for a given value of the parameter as
\begin{eqnarray}\label{illuminationmode}
\bar{E}(\vec{r},p)=2\sqrt{N} u_{0}(\vec{r},p),
\end{eqnarray}
where $N$ refers to the total number of photons detected in the mean
field during the integration time of the detector. $N$ is assumed to
be independent of $p$, as stated previously. $u_{0}(\vec{r},p)$ is
the $p$-dependent transverse distribution of the mean field, complex
in the general case, and normalized to $1$ (its square modulus
integrated over the transverse plane equals 1). The local mean
photon number detected during the same time interval is
\begin{eqnarray}
\bar{n}(\vec{r},p)=\frac{\bar{E}^{*}(\vec{r},p)\bar{E}(\vec{r},p)}{4}=N|u_{0}(\vec{r},p)|^{2}.
\end{eqnarray}

{\it Intensity measurements.} - We first assume that $p$ is
determined by processing the information contained in the
measurement of the local intensity, i.e. local number of photons.
The best possible local intensity detection device would consist in
a set of indexed pixels paving the entire transverse plane, in the
limit when their spatial extension approaches $0$. Let
$\textbf{n}=\left[n_1,...,n_k,...\right]$ be one measurement of the
photon distribution with such a hypothetical detector, where $n_k$
corresponds to the number of photons detected on pixel $k$. Because
of the noise present in the light, the sample $\textbf{n}$ differs
from its statistical mean value
$\bar{\textbf{n}}(p)=\left[\bar{n}_1(p),...,\bar{n}_k(p),...\right]$.
Let $L(\textbf{n}|p)$ be the likelihood of its observation. Note
that $\textbf{n}$ corresponds to a single measurement and hence does
not explicitly depend on $p$, contrarily to the average on all the
possible realizations.

The achievable precision on the estimation of $p$ is limited by the
Cramer-Rao Bound (CRB). More precisely, the variance of any unbiased
estimator of $p$ is necessarily greater than $\sigma_{min}^2=1/I_F$,
where the Fisher information $I_{F}$ is given, when the actual value
of $p$ is $0$, by \cite{Refregier2002}:
\begin{equation}\label{Fisher}
I_F=-\int \left. \frac{\partial^2}{\partial p^2}l(
\textbf{n}|p)\right|_{p=0} L(\textbf{n}|0)d\textbf{n},
\end{equation}
where we have introduced the log-likelihood
$l(\textbf{n}|p)=\ln{L(\textbf{n}|p)}$. The integration spans
continuously over all possible photon distributions that can be
detected when the parameter value is $p=0$. This information is thus
highly dependent on the noise statistics.

Let us first assume that the illumination is coherent, and therefore
that the local intensity noise is Poissonian. The probability of
measuring $n_k$ photons on pixel $k$, when the parameter equals $p$
is given by
\begin{equation}\label{proba}
P_{k,p}(n_k)=\frac{\bar{n}_k(p)^{n_k}}{n_k!}e^{-\bar{n}_k(p)}.
\end{equation}
Restricting our analysis to spatially uncorrelated beams, the
likelihood $L(\textbf{n}|p)$ simply corresponds to the product of
all local probabilities given in Eq.~\ref{proba}. Then, taking the
limit of infinitely small pixels and using Eq.~\ref{Fisher}, one can
show that the Fisher information equals
\begin{equation}
I_F^{Poisson}=\int\left[\frac{\bar{n}'(\vec{r},0)^2}{\bar{n}(\vec{r},0)}-
\bar{n}''(\vec{r},0)\right]d^2r,
\end{equation}
where the $'$ denotes a derivative relative to $p$. Using
Eq.~\ref{illuminationmode}, one finally finds that
\begin{equation}
I_F^{Poisson}=\frac{4N}{a^2},
\end{equation}
where $a$ is a global positive parameter characterizing the
variation of the image intensity with $p$, defined by
\begin{equation}\label{defa}
\frac{1}{a^2}=\int\left[\frac{\partial}{\partial
p}|u_0(\vec{r},p)|\right]_{p=0}^2d^2r.
\end{equation}

The smallest value of $p$ that can be distinguished from the shot
noise - i.e. corresponding to a signal to noise ratio (SNR) equal to
1 -, whatever the algorithm used to determine it from the local
intensity measurements, provided that it gives an unbiased
estimation of $p$, is finally greater than $a/2 \sqrt{N}$. This
value sets the standard quantum noise limit for intensity
measurements of $p$, imposed by the random time arrival of photons
on the detector. It is inversely proportional to the square root of
the number of photons, as expected, and is related, through the
dependence with $a$, to the modification of the mean intensity
profile with $p$.

We now consider a non-classical illumination, still with identical
mean intensity, but with local sub-Poissonian quantum fluctuations
described by a noise variance $\sigma_P^2~<~1$ (assumed to be the
same over the entire transverse plane). One can show that the CRB
leads to
\begin{eqnarray}\label{CRB2}
p_{min}^{sub-Poisson}\ge\frac{a\sigma_P}{2\sqrt{N}}.
\end{eqnarray}

As we have already noticed, the limit given by Eq.~\ref{CRB2} is
valid for any measurement strategy. Nevertheless, a practical way
enabling us to reach such an absolute limit remains to be found.
This is what is presented in the next paragraph.

\begin{figure}[h]
\begin{center}
\includegraphics[width=4cm]{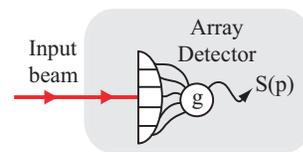}
\caption{Array detector as an optimal intensity detection. $g$
refers to the pixel gain distribution. $S(p)$ is the p-dependent
processed signal.} \label{Figarray}
\end{center}
\end{figure}

Let us assume that an image processor calculates a given linear
combination of the local intensity values recorded by the pixels of
an array detector, as represented in Fig.~\ref{Figarray}.
Assuming that the pixels are small compared to the characteristic
variation length of the image, the mean value $\bar{S}(p)$ of the
computed signal can be written as an integral over the transverse
plane as follows
\begin{eqnarray}\label{array}
\bar{S}(p)=\int g(\vec{r})\bar{n}(\vec{r},p)d^2r,
\end{eqnarray}
where $g(\vec{r})$ is the local gain on the pixel localized at
position $\vec{r}$, which can be positive or negative. Assuming that
$p$ is small, $|u_{0}(\vec{r},p)|$ can be expanded at first order
into
\begin{eqnarray}
|u_{0}(\vec{r},p)|=|u_{0}(\vec{r},0)|+\frac{p}{a}u_{I}(\vec{r}),
\end{eqnarray}
where $a$ has been defined in Eq.~\ref{defa}, and $u_I$ is a
transverse function normalized to $1$, already introduced in a
particular case in reference \cite{Delaubert2006}. The gains are
chosen such as $\bar{S}(p=0)=0$ (difference measurement), so that
$\bar{S}(p)$ is given, at first order, by
\begin{eqnarray}
\bar{S}(p)=\frac{2Np}{a}\int
g(\vec{r})|u_{0}(\vec{r},0)|u_{I}(\vec{r})d^2r,
\end{eqnarray}
For a coherent illumination, because the local quantum fluctuations
are uncorrelated, the noise variance $\Delta S^{2}$ on  on
$\bar{S}(p)$ is equal to the shot noise on each pixel weighted by
$g^{2}(\vec{r})$ \cite{Treps2005}. Moreover, as $p$ is small, the
noise is independent of $p$ at the first order, and we get
\begin{eqnarray}
\Delta S^{2}=N\int g^{2}(\vec{r})|u_{0}(\vec{r},0)|^{2}d^2r.
\end{eqnarray}

It is then possible to optimize the gain factor $g(\vec{r})$ in
order to get the highest possible SNR defined by
SNR$=\bar{S}(p)^{2}/\Delta S^{2}$. Using Cauchy-Schwartz inequality,
one can show that the highest SNR value for the present measurement
strategy is given by SNR$= 4Np^{2}/a^{2}$, and is obtained for an
optimal value of the gain distribution given by
\begin{eqnarray}
g_{opt}(\vec{r})=\beta \frac{u_I(\vec{r})}{ |u_{0}(\vec{r},0)|},
\end{eqnarray}
where $\beta$ is an arbitrary constant. The minimum measurable value
of $p$ - corresponding to comparable signal and noise, i.e. SNR$=1$
- is given by $p_{min}= a/2 \sqrt{N}$, which is precisely equal to
the CRB for classical illumination. The present measurement strategy
is therefore optimal as it allows to reach the CRB for small values
of the parameter $p$, with the certainty that no other measurement
strategy can do better. Note that we have not proven that it is the
unique way to reach the CRB.

We can even extend this result to the use of non classical light.
Indeed, using a bimodal field composed of a bright mode $u_{0}$ in a
coherent state carrying the mean field, and a squeezed vacuum mode
in the mode $u_{I}$, the detected noise power is modified into
%
\begin{eqnarray}
\Delta S^{2}=N\sigma_{P}^{2},
\end{eqnarray}
when the noise variance on the amplitude quadrature of the mode
$u_{I}$ is given by $\sigma_{P}^{2}$ \cite{Treps2005}. Note that the
use of a locally squeezed beam would not provide any improvement as
$u_I$ is the only mode contributing to the measurement noise,
referred to as the {\it noise-mode of detection}
\cite{Delaubert2006,Treps2005}. The minimum measurable $p$ value is
in this case
\begin{eqnarray}
p_{min}= \frac{a\sigma_{P}}{2 \sqrt{N}}.
\end{eqnarray}
We have thus found a way to reach the bound, i.e. the minimum
accessible $p$-value given in Eq.\ref{CRB2}. Moreover, our scheme
requires minimum quantum resources, namely a bimodal field with
squeezing in only one mode.


{\it Field measurements.} - We now assume that the information about
$p$ is extracted from the knowledge of the local complex field, i.e.
local amplitude and phase, that can be obtained by interferometric
techniques. Similarly to the previous section, the best possible
detection would here access the local complex field on k-indexed
areas paving the entire transverse plane, in the limit when their
spatial extension approaches $0$. Let
$\textbf{E}=\left[E_1,...,E_k,...\right]$ be a single measurement of
the field distribution, hence independent of $p$, where $E_k$
corresponds to the complex field detected on area $k$. Again,
because of the noise present in the light, the sample $\textbf{E}$
differs from its statistical mean value
$\bar{\textbf{E}}(p)=\left[\bar{E}_1(p),...,\bar{E}_k(p),...\right]$.
Let $L(\textbf{E}|p)$ be the likelihood of its observation.

A bound to the maximal precision on $p$ can again be calculated from
the CRB, which, for a field measurement, is the inverse of the
following Fisher information
\begin{eqnarray}\label{IF}
I_{F}=-\int\left[\frac{\partial^{2}}{\partial
p^{2}}l(\textbf{E}|p)\right]_{p=0}L(\textbf{E}|0)d\textbf{E},
\end{eqnarray}

We assume that the local field fluctuations in the transverse plane
can be described by a Gaussian probability density function
independent of the mean field. Moreover, we consider a classical or
non classical illumination whose amplitude and phase quadrature
fluctuations are described by the noise variances $\sigma_P^2$ and
$\sigma_Q^2$, respectively. These factors neither depend on
$\vec{r}$ nor on $p$, as we assume the fluctuations to be
homogeneous and independent of the parameter $p$.

The probability to measure a field given by $E_k=P_k+iQ_k$ on area
$k$, where $P_k$ and $Q_k$ correspond to the local field
quadratures, is, for a parameter value $p$
\begin{eqnarray}\label{PE}
\mathcal{P}_{k,p}(E_k)=\frac{1}{2\pi \sigma_P\sigma_Q}
e^{-\left[\frac{\left(P_k-\bar{P}_k(p)\right)^2}{2\sigma_P^2}+\frac{\left(Q_k-\bar{Q}_k(p)\right)^2}{2\sigma_Q^2}\right]},
\end{eqnarray}
where $\bar{P}_k(p)$ and $\bar{Q}_k(p)$ are the local quadratures
statistical averages, satisfying:
$\bar{E}_k(p)=\bar{P}_k(p)+i\bar{Q}_k(p)$. Without loss of
generality, we define the orientation of the Fresnel diagram
relative to the phase of the mean field, i.e.
$\bar{P}(\vec{r},p)=2\sqrt{N}u_0(\vec{r},p)$ and
$\bar{Q}(\vec{r},p)=0$, taking the limit of infinitely small
detection areas. Assuming no spatial correlations in the field
fluctuations, one can show that the Fisher information is given by
\begin{eqnarray}\label{IF1}
I_{F}^{Gauss}&=&\int\frac{1}{\sigma_{P}^{2}}\left[\frac{\partial
\bar{P}(\vec{r},p)}{\partial p}\right]^{2}_{p=0}d^{2}r.
\end{eqnarray}
We now introduce $b$, a second global positive parameter,
characterizing the variation of the image field with $p$
\begin{equation}\label{defb}
\frac{1}{b^2}=\int\left[\frac{\partial}{\partial
p}u_0(\vec{r},p)\right]_{p=0}^2d^2r.
\end{equation}
Using Eq.\ref{illuminationmode} and \ref{defb}, the Fisher
information simplifies into
\begin{eqnarray}
I_{F}^{Gauss}=\frac{4N}{b^{2}\sigma_{P}^{2}}.
\end{eqnarray}
The smallest value of $p$ that can be distinguished from the quantum
noise using a field detection of the optical beam is finally greater
than the CRB:
\begin{eqnarray}
p_{min}^{Gauss}\ge\frac{b\sigma_{P}}{2\sqrt{N}}.
\end{eqnarray}


Again, we can propose an experimental scheme enabling to reach this
limit in the case of a small parameter $p$, and for which the mean
value of the complex electric field can be written at first order
\begin{eqnarray}\label{Esignal}
\bar{E}(\vec{r},p)=2\sqrt{N}\left[
u_{0}(\vec{r},0)+\frac{p}{b}u_{E}(\vec{r})\right],
\end{eqnarray}
where $b$ has been introduced in Eq.\ref{defb}, and where
$u_E(\vec{r})$ is a transverse function normalized to $1$.

\begin{figure}[h]
\begin{center}
\includegraphics[width=5cm]{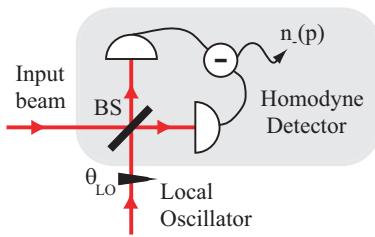}
\caption{Balanced homodyne detection as an optimal field detection.
$n_-(p)$ is the p-dependent signal. $\phi_{LO}$ is the local
oscillator phase. BS : $50/50$ beamsplitter.} \label{homodyne}
\end{center}
\end{figure}

Let us consider a balanced homodyne detection, as represented in
Fig. \ref{homodyne}, with a local oscillator (LO) chosen to be
defined by the following complex field operator
\begin{eqnarray}\label{ELO}
\bar{E}_{LO}(\vec{r})=2\sqrt{N_{LO}} u_{E}(\vec{r})e^{i\theta_{LO}},
\end{eqnarray}
where $N_{LO}$ corresponds to the number of photons detected in the
entire LO beam during the integration time.
The LO is much more intense than the image, i.e. $N_{LO}\gg N$.
$\theta_{LO}$ is the LO phase. The mean intensity difference $n_{-}$
between the photocurrents of the two detectors
is given in terms of number of photons by
\begin{eqnarray}\label{hd}
\bar{n}_-(p)=\frac{1}{4}\int\left[
\bar{E}^*_{LO}(\vec{r})\bar{E}(\vec{r},p)+\bar{E}^*(\vec{r},p)\bar{E}_{LO}(\vec{r})\right]d^2r.
\end{eqnarray}
Note the similarity with Eq.\ref{array}, as incident field amplitude
and LO field play here identical roles of incident intensity and
electronic gain, respectively. Though detectors do not resolve the
spatial distribution of the beams and no processing of the spatial
information is made, the balanced homodyne technique directly
provides an "analog" computation of the quantity of interest.


When the LO phase is tuned to the maximum of the $p$-dependent term,
the homodyne signal becomes
%
%
\begin{eqnarray}\label{nminus}
\bar{n}_{-}=2\sqrt{NN_{LO}}\frac{p}{b}.
\end{eqnarray}
For coherent illumination, the noise power on the homodyne signal
corresponds to $N_{LO}$, i.e. to the shot noise of the LO. The SNR
of the homodyne measurement is thus given by SNR$=4Np^{2}/b^{2}$.
The minimum measurable value of $p$ - corresponding to a SNR of $1$
- with homodyne detection is given by
\begin{eqnarray}
p_{min}=\frac{b}{2\sqrt{N}}.
\end{eqnarray}
Moreover, when the component of the image selected by the LO is in a
non classical state , i.e. allowing a squeezed vacuum $u_{E}$ mode
with a noise variance $\sigma_{P}^{2}$ on the amplitude quadrature
within the incoming beam, we get
\begin{eqnarray}
p_{min}=\frac{b\sigma_{P}}{2\sqrt{N}}.
\end{eqnarray}
This result corresponds exactly to the CRB calculated for amplitude
measurements of $p$. The homodyne detection scheme with the
appropriate LO shape and phase is therefore an optimal field
detection of $p$. Again, it uses minimal resources as only one
source of squeezed light is needed to reach the non classical CRB.


{\it Comparison and conclusion}. - We have presented two efficient -
i.e. reaching the associated CRB - signal processing techniques for
the extraction of information contained in an image. We can show
that the CRB for field measurements, in which all quadratures can be
accessed, is smaller than the one for intensity measurements, i.e.
$a\ge b$. Yet, both schemes are useful: the intensity scheme is
interesting since it is not restricted to monochromatic light,
whereas the amplitude scheme is useful since it does not require
pixellized detectors.

Let us finally note that this work provides limits which are valid
for a shot noise limited light of any shape. However it is not valid
so far for any kind of non-classical light, as we have restricted
our analysis to homogeneous squeezed states. In a forthcoming
publication, we will study the situation of quantum fields with
quantum spatial correlations in the transverse plane and will
investigate for the corresponding modifications of the CRB.

{\it Acknowledgements}. We would like to acknowledge the support of
the Australian Research Council scheme for Centre of Excellence.
Laboratoire Kastler Brossel, of the Ecole Normale Sup\'erieure and
Universit\'e Pierre et Marie Curie, is associated with the CNRS.

\end{document}